\date{}
\documentclass[showpacs]{revtex4}
\usepackage{graphicx,psfrag,amsmath,amssymb,amsfonts,bbm,latexsym,color,dcolumn,epsf}

\begin{document}

\title{Decoherence in composite quantum open systems: the effectiveness of unstable
degrees of freedom}

\author{Fernando C. Lombardo \footnote{Electronic address: lombardo@df.uba.ar}}
\author{Paula I. Villar \footnote{Electronic address: paula@df.uba.ar}}
\affiliation{Departamento de F\'\i sica {\it Juan Jos\'e Giambiagi}, FCEyN UBA,
Facultad de Ciencias Exactas y Naturales, Ciudad Universitaria,
Pabell\' on I, 1428 Buenos Aires, Argentina}

\date{today}

\begin{abstract}
The effect induced by an environment on a composite quantum system is studied. The model considers the 
composite system as comprised by a subsystem A coupled to a subsystem B which is also coupled to an 
external environment. We study all possible four combinations of subsystems A and B made up with a 
harmonic oscillator and an upside down oscillator. We analyzed the decoherence suffered by subsystem A  
due to an effective environment composed by subsystem B and the external reservoir. In all the cases we 
found that subsystem A decoheres even though it interacts with the 
environment only through its sole coupling to B. However, the effectiveness of the diffusion depends on 
the unstable nature of subsystem A and B. Therefore, the role of this degree of freedom in the 
effective environment is analyzed in detail. 
\end{abstract}


\maketitle

\newcommand{\beq}{\begin{equation}}
\newcommand{\eeq}{\end{equation}}
\newcommand{\beqa}{\begin{eqnarray}}
\newcommand{\eeqa}{\end{eqnarray}}
\newcommand{\beqas}{\begin{eqnarray*}}
\newcommand{\eeqas}{\end{eqnarray*}}

\section{Introduction}

Decoherence is the process by which most pure states evolve into
mixtures due to the interaction with an environment
\cite{zurek-gral}. The very notion of a quantum open system implies
the appearance of dissipation and decoherence as an ubiquitous
phenomena and plays important roles in different branches of physics
\cite{more} (from quantum field theory \cite{nos}, many body and molecular
physics to theory of quantum information), biology and chemistry.
Oftentimes, a large system, consisting of two or a few subsystems
(degrees of freedom) interacting with their environment (thermal
bath comprising a large number of degrees of freedom), can be
adequately described as a composite system. Examples include
electron transfer in solution \cite{prezhdo}, a large biological
molecule, vibrational relaxation of molecules in solution, excitons
in semiconductors coupled to acoustic or optical phonon modes.
Quantum processes in condensed phases are usually studied by
focusing on a small subset of degrees of freedom and considering the
other degrees of freedom as a bath.

Another interesting aspect is referred to the quantum to classical
transition. The emergence of classicality is a typical
consequence of having the quantum system in direct interaction with
the external world. In fact, decoherence is the main ingredient in
order to find classicality. The interaction between system and
environment induces a preferred basis which is stable against this
interaction and becomes a classical basis in the Hilbert space of
the coupled system. Preferred pointer states are resilient to the
entangling interaction with the bath. This ``einselection''
(environment induced superselection) of the preferred set of
resilient pointer states is the essence of the environment. It is
accepted that a rapid loss of coherence caused by the coupling with
the environment is at the root of the non-observation of quantum
superpositions of macroscopically distinct quantum states
\cite{JPP-LH}.

In this article, we analyze the decoherence induced in a composite
quantum system, in which an observer can distinguish between two
different subsystems, one of them coupled to an external
environment. Our composite system is composed by a subsystem A coupled
to a subsystem B which is also bilinearly coupled to an external environment $\cal{E}$.
The coupling to this external environment is only through subsystem B.
Subsystem A remains isolated from $\cal{E}$ but for the information delivered
by B through a bilinearly coupling between subsystems A and B. We will consider
the thermal bath to be at high temperature and will work in the underdamped limit.

In order to investigate this problem we mainly consider a simple
model where subsystem A is represented by a harmonic oscillator and
subsystem B is an upside-down one. The main motivation for studying
this model is twofold. On the one hand, it is of interest to deepen
and enlighten previous analysis of decoherence induced by chaotic
environments. The upside-down oscillator has  recently been used to
model a chaotic environment which induces decoherence on the system
\cite{robin}. Even though it is an oversimplified model for a
chaotic environment, it displays exponential sensitivity to
perturbations, which is crucial in order to analyze chaotic
evolutions. On the other hand, we want to emphasize that isolation
from a chaotic environment is difficult, as has been noted in
\cite{robin}. Moreover, it is even harder to isolate a system from a
chaotic environment than from the many harmonic oscillators of the
quantum Brownian motion environment. In this context, we shall
consider two different cases. Firstly, the case where the chaotic degree of
freedom is part of the environment (i.e. an unstable system B) and
is directly coupled to an external reservoir $\cal{E}$ and to another
subsystem A with different bare frequency. Secondly, the case where
subsystem A is
unstable and directly coupled to a harmonic oscillator (subsystem B)
which is also coupled to an external bath $\cal{E}$. These are the
extension of previous works done in \cite{robin} and \cite{elze,diana}
for the first and second case, respectively. In both situations, we
will estimate the decoherence time, which is the usual scale after
which classicality emerges and is different for each case. We will
show the dependency of these times with the parameters of the
model.

The analysis is completed by the inclusion of the other two
different possibilities for the quantum composite system, i.e., a
composite system constituted by a subsystem A coupled to subsystem
B, both harmonic oscillators, and a composite system formed by
subsystem A coupled to subsystem B, both inverted oscillators. As in
the other two cases mentioned above, subsystem B is also coupled to
an external reservoir $\cal{E}$. All in all, we have four different
composite systems to analyze.

For every and each situation, we study the dynamics of the subsystem
A. Not only did we study the influence of ''its" environment (formed
by subsystem B and $\cal{E}$) at high temperature but also
in the absence of the external reservoir $\cal{E}$. In every case,
we conclude that decoherence is faster in the case in which
subsystem A is unstable. However, different conclusions can be
arrived at when the subsystem A is a harmonic oscillator. All the
cases studied in this paper have different decoherence time-scales
associated, depending not only on the external temperature, but also
on the type of subsystem one is considering in turn. Each case
develops a different dynamics, being possible, sometimes, to
find a quantum open system described using mixed quantum-classical
dynamics \cite{kapral} (part of the composite system completely
decohered while others didn't).

In previous articles, a different composite system has been considered. For
example, in Ref.\cite{kapral1}, subsystem A is taken to be a two-level system
which is bilinearly coupled to a single harmonic oscillator B-subsystem; which 
is also coupled to an Ohmic (or super-Ohmic) set of infinite
harmonic oscillators. Authors have shown that subsystem B losses coherence
more rapidly than subsystem A, which maintains coherence for longer periods of time.
This two-level composite system was also studied in \cite{anu}
 in order to look at the exact solutions
for the dynamics of the reduced density matrix of the composite
system AB. Even though the composite system presented in this paper
might appear to be similar to the one in \cite{kapral1,anu}, it has
a different dynamics and therefore is of interest to investigate it
separately. This article is an extended version of Ref. \cite{prabr}, 
where we firstly presented our model and numerically evaluated the decoherence 
times in every case. Here, we are showing a complete analytical development of the influence 
functional method for the composite system (what is not included in \cite{prabr}). From the 
influence action we evaluate the diffusive corrections to the master equation, and 
we complete the analysis about decoherence times, providing analytical and numerical 
estimations of the decoherence times, based on the inverted oscillators' dynamics in the 
phase space.

The paper is organized as follows. In Section II we present our
model and evaluate the influence \mbox{functional} for each of the
considered cases, in order to compute, in Section III, the diffusion
coefficient of the master equation for the subsystem A. We \mbox{evaluate}
diffusion analytically. Section IV is devoted to the analysis of the
decoherence process. This is done by means of the decoherence
factor. Section V contains our final remarks and in the Appendix we
include details of the calculations.

\section{The model: Composite quantum system in an external environment}

\subsection{General formulation}

We consider a AB${\cal E}$ quantum system consisting of three coupled
subsystems: subsystem A is coupled directly to subsystem B,
while subsystem B is in direct contact with an external
environment ${\cal E}$. The total AB${\cal E}$ classical action is,

\beq
S[x,q,Q] = S_{\rm A}[x] + S_{\rm B}[q] + S_{\rm {\cal E}}[Q] +
S_{\rm AB}[x,q] + S_{B \cal E}[q,Q].
\eeq

In the spirit of the quantum Brownian paradigm, the environment is
taken to be a set of $N$ independent harmonic oscillators with
frequencies $\tilde{\omega }_n$, masses $m_n$ and coordinates and
conjugate momenta
$(\hat{Q},\hat{P})=(\hat{Q}_1,...,\hat{Q}_N,\hat{P}_1,...,\hat{P}_N)$
so that the classical action is

\beq S_{\rm {\cal E}}[Q] = \int_0^t ds \sum_n {m_m\over{2}} ({\dot Q}_n^2 -
\tilde{\omega}_{n}^2 Q_n^2). \eeq

Subsystem B consists of a single oscillator (upside-down or
harmonic, depending on the case considered) with bare mass
$M_{\rm B}$, frequency $\Omega$ and coordinate operator $\hat{q}$,

\beq
S_{\rm B}(x) = \int_0^t ds {M_{\rm B}\over{2}} (\dot q^2 \pm
\Omega^2 q^2). \eeq

The interaction between subsystem B and the thermal environment is assumed to be
bilinear,

\beq S_{\rm B {\cal E}}^{\rm int} = \int_0^t ds \sum_n c_n q(s)
Q_n(s), \eeq
where $c_n$ is the coupling constant to the $n\rm{th}$ oscillator.
The environment is characterized by the spectral density
$I_{\cal E} \equiv  \pi \sum_n \frac{c_n^2}{2 m_n
\tilde{\omega}_n} \delta (\tilde{\omega}-\tilde{\omega}_n)$.
For simplicity, we assume an Ohmic environment,
with the following spectral density $I_{{\cal E}}(\omega ) = 2 M \gamma_0 \tilde{\omega}
e^{{-{\tilde{\omega}^2\over{\Lambda^2}}}}$, where $\Lambda$ is a
physical cutoff, related to the maximum frequency present in the environment.

Finally, we consider subsystem A consisting of a single
oscillator (again, this oscillator can be an upside-down or harmonic one)
with coordinate operator $\hat{x}$ whose classical action is
 \beq
S_{\rm A}[x] = \int_0^t ds {M_{\rm A}\over{2}} (\dot x^2 \pm
\omega^2 x^2). \eeq

We suppose that subsystem A is bilinearly coupled to subsystem B
by the interaction term \beq S_{\rm AB}^{\rm int} = - \lambda \int_0^t ds
x(s) q(s).\label{intAB} \eeq

The dynamical properties of interest can be computed from the
density matrix of the system at time $t$. The complete density
matrix may be written in integral form in terms of the total
propagator (we are setting the initial time $t_0 = 0$)

\beq \hat{K}(x,q,Q;t|x_0,q_0,Q_0;0) \equiv \hat{K}(t|0) 
= <x q Q\vert
\exp(-i \hat{H}t / \hbar)\vert x_0 q_0 Q_0> \nonumber \eeq
as

\beq
\hat{\rho}(x,q,Q,x\prime,q\prime,Q\prime) =\int
dx_0 dx_0\prime dq_0 dq_0\prime dQ_0 dQ_0\prime \hat{K}(t|0)
\hat{\rho}(0)
\hat{K}^\ast(t|0).\eeq

We are primarily interested in the dynamics of the composite AB-system
under the influence of the external environment $\cal E$.
In such case, the relevant quantity to analyze is the reduced density
matrix $\hat{\rho}_{\rm r}$,
obtained by integrating out the environmental degrees of freedom.
Such a reduction is especially correct if the characteristic time
scale of the environment (which essentially is $1/\Lambda$) is much shorter
than those for the subsystem A and subsystem B. As is usually done, we
assume a
factorized initial condition between the composite system AB and the
environment $\cal E$, \beq \hat{\rho}(x_0,x_0\prime,q_0,q_0
\prime, Q_0,Q_0\prime;0) = \hat{\rho}_{\rm AB}(x_0,x_0
\prime,q_0,q_0\prime;0) \hat{\rho}_{\cal
E}(Q_0,Q_0\prime;0), \eeq and the external environment initially
in thermal equilibrium at temperature T.

In this way we can write the integral form of the reduced density
matrix at time $t$ as \beq
\hat{\rho}_{\rm r}(x,x\prime,q,q\prime,t)= \int
dx_0 dx_0 \prime dq_0 dq_0\prime \hat{J}_{\rm r}(x,x\prime,q,q\prime;t
\vert x_0,x_0\prime,q_0,q_0\prime;0) 
\hat{\rho}_{\rm AB}(x_0,x_0 \prime,q_0,q_0\prime;0), \eeq
where the reduced time evolution operator $\hat{J}_{\rm r}$ is \beqa
\hat{J}_{\rm r}(x,x\prime,q,q\prime;t \vert
x_0,x_0\prime,q_0,q_0\prime;0) &=& \int
dQ_0dQ_0\prime \hat{K}(x,q,Q;t\vert x_0,q_0,Q_0;0) \rho_{\cal
E}(Q_0,Q_0\prime,0)\nonumber \\
&\times & \hat{K^{\ast}}(x\prime,q\prime,Q\prime;t\vert
x_0\prime,q_0\prime,Q_0\prime;0).\eeqa

Given the initial conditions mentioned above, this expression for
the reduced density matrix specifies a non-Markovian time evolution
since the solution at time $t$ depends on its past history. In the
following we will use the influence functional method for deriving
the master equation. This method provides us a way to obtain a
functional representation of the evolution operator $\hat{J}_{\rm r}$ for
the reduced density matrix.

\subsection{Influence functional method}

The formulation of the reduced density matrix in terms of an
influence functional is widely
discussed in the literature \cite{fey,Hu,Grabert}. In the present paper, we
will extend it to the composite AB-system, comprising two
oscillators (harmonic or inverted). In the general setting,
the evolution operator $\hat{J}$ for the full density matrix $\hat\rho$
is $\hat{\rho}(t)=\hat{J}(t,0)\hat{\rho}(0)$, where
\beqa
\hat{J}(x_{\rm f},q_{\rm f},Q_{\rm f},x_{\rm f}\prime,q_{\rm f}\prime,Q_{\rm f}'
\vert x_0,q_0,Q_0,x_0',q_0',Q_0')
&=&  \int_{x_0}^{x_{\rm f}}{\cal D}x\int_{q_0}^{q_{\rm f}}{\cal
D}q\int_{Q_0}^{Q_{\rm f}}{\cal D}Q e^{{i\over{\hbar}}S(x,q,Q)} \nonumber \\
&\times & \int_{x_0'}^{x_{\rm f}'}{\cal D}x'\int_{q_0'}^{q_{\rm f}'}{\cal D}q'
\int_{Q_0'}^{Q_{\rm f}'}{\cal D}Q' e^{-{i\over{\hbar}}S(x',q',Q')}
.\eeqa The path integrals are over all histories
compatible with boundary conditions. As was mentioned in the Section
above, our primary interest is in the effect of the external
environment on our composite system AB. Therefore, we  need the reduced
density matrix for the AB system, defined as \beq
\rho_{\rm r}(x,x',q,q') 
=  \int_{-\infty}^{+\infty}dQ \int_{-\infty}^{+\infty}dQ' \rho
(x,q,Q\vert x',q',Q') \delta (Q - Q'), \eeq and the evolution in
time is given by the reduced evolution operator $\hat{J}_{\rm r}$

\beq
\rho_{\rm r}(x,x',q,q';t)=\int_{-\infty}^{+\infty}
\int_{-\infty}^{+\infty}dx_0 dx_0' dq_0 dq_0' ~ J_{\rm
r}(t\vert 0) 
\rho_{\rm AB}(x_0,x_0',q_0,q_0';0). \eeq

Assuming total separable initial conditions as mentioned above, the
reduced propagator is

\beqa
J_{\rm r}(x_{\rm f},x_{\rm f}',q_{\rm f},q_{\rm f}';t
\vert x_0,x_0',q_0,q_0';0) 
&=&  \int_{x_0}^{x_{\rm f}}{\cal D}x \int_{x_0'}^{x_{\rm f}'}{\cal
D}x' \int_{q_0}^{q_{\rm f}}{\cal D}q \int_{q_0'}^{q_{\rm f}'}{\cal D}q'
e^{{i\over{\hbar}}(S_{\rm A}(x) - S_{\rm A}(x'))} \nonumber \\ &\times &
e^{{i\over{\hbar}}( S_{\rm B}(q) - S_{\rm B}(q'))} e^{{i\over{\hbar}}( S_{\rm AB}(x,q)
- S_{\rm AB}(x',q'))} ~ F(x,x',q,q'),
\eeqa
where $F(x,x',q,q')$ is the Feynmann-Vernon influence functional
\cite{fey} (now for the composite system) given by

\begin{eqnarray}
F(x,x',q,q') 
&=&\int_{-\infty}^{+\infty} dQ_{\rm f}
\int_{-\infty}^{+\infty}dQ_0\int_{-\infty}^{+\infty} dQ_{\rm f}'
\int_{Q_0}^{Q_{\rm f}}{\cal D}Q\int_{Q_0'}^{Q_{\rm f}'}{\cal D}Q'  \nonumber \\
&\times &e^{{i\over{\hbar}}(S_{{\cal E}}(Q) + S_{\rm B{\cal E}}(q,Q) -
S_{{\cal E}}(Q') - S_{\rm B{\cal E}}(q',Q'))} \ \rho_{{\cal E}}(Q_0,Q_0')  \nonumber \\
&\equiv & e^{{i\over{\hbar}} \delta A(x,x',q,q')},\end{eqnarray}
where $\delta A(x,x',q,q')$ is the influence action for the
AB composite system. Thus, we can define
$A(x,x',q,q')$ as the coarse graining effective action:
$A(x,x',q,q')= S_{\rm A}(x) - S_{\rm A}(x') + S_{\rm B}(q)
- S_{\rm B}(q')+ S_{\rm AB}(x,q)-
S_{\rm AB}(x',q') + \delta A(x,x',q,q')$.

It is important to note that in our model, the subsystem A is
not directly coupled to the environment. Consequently, the influence
functional is the well known influence functional  $\delta A$ for a
bath of harmonic oscillators found in the literature (and only a
function of $q$ and $q'$) \cite{Hu}

\beq \delta A(q,q') =  -2
\int_0^t ds \int_0^s ds' \Delta q(s) ~\eta (s - s') ~\Sigma q(s')
+ i \int_0^t ds \int_0^s ds' \Delta q(s) ~\nu (s -
s') ~\Delta q(s') \label{altaT}\eeq with

\beq  \Delta q(s) = q(s) - q'(s) ~;~
 \Sigma q(s) = \frac{1}{2} (q(s) +
q'(s)). \label{Deltaq} \eeq The kernels $\eta$ and $\nu$ (dissipation and noise,
respectively) are in general non-local and are defined as
$\eta (s) = \frac{d}{ds}\gamma (s)$, 
with
$$ \gamma (s) = \int_0^\infty d\omega \frac{I_{{\cal E}}(\omega )}{\omega} \cos \omega s,$$
and
$$\nu (s) = \int_0^\infty d\omega ~I_{{\cal E}}(\omega ) \coth \frac{\beta \hbar\omega}{2}
\cos \omega s,$$ up to second order in the coupling constant with the external 
environment. In the high temperature limit, these kernels are
proportional to $\nu \sim 2 M \gamma_0k_B T \delta (s)/\hbar$  and
$\eta \sim  M \gamma_0 \dot{\delta}(s)$ \cite{Hu,habib}. As we are
working in the underdamped and high temperature limit (which means
$k_BT\gg\hbar \omega$ and $\gamma_0 \ll \Omega$ but leaves $\gamma_0k_BT$
unrestricted), we can use the latter expressions. Therefore, if we
evaluate Eq.(\ref{altaT}), we have \beq \delta A
(q,q\prime) \simeq - 2 M_{\rm B} \gamma_0 \int_0^t ds \Delta q(s)
\dot{\Sigma} q(s)+ i \frac{2M_{\rm B}\gamma_0k_B
T}{\hbar} \int_0^t (\Delta q(s))^2 ds. \label{deltaA} \eeq

Consequently, after integrating the bath of harmonic oscillators, we can
write the influence functional $F(q,q\prime)$, in the high temperature limit
as

\beq F(q,q\prime) = e^{-\frac{i2 M_{\rm B} \gamma_0}{\hbar}
\int_0^t ds \Delta q(s)
\dot {\Sigma} q(s) } e^{\frac{-2M_{\rm B}\gamma_0k_B T}{\hbar^2} \int_0^t (\Delta q(s))^2
ds}, \eeq
and therefore the reduced density matrix takes the form \beqa
\rho_{\rm r}(x,x',q,q') 
& = & \int_{-\infty}^{\infty} dx_0 dx_0\prime
\int_{-\infty}^{\infty}dq_0 dq_0\prime \int_{-\infty}^{\infty}dq_{\rm f} dq_{\rm f}\prime 
\int_{q_0}^{q_{\rm f}} {\cal D} q \int_{q_0\prime}^{q_{\rm f}\prime} {\cal D}
q\prime \ e^{\frac{i}{\hbar}(S_{\rm B}(q)-S_{\rm B}(q\prime))}\nonumber \\ &\times &
\int_{x_0}^{x_{\rm f}} {\cal D} x
 \int_{x_0\prime}^{x_{\rm f}\prime} {\cal D} x\prime \
e^{\frac{i}{\hbar}(S_{\rm A}(x)-S_{\rm A}(x\prime))}
e^{\frac{i}{\hbar}\big(S_{\rm AB}(x,q)-S_{\rm AB}(x\prime,q\prime)+ \delta
A(q,q\prime)\big)}. \eeqa

At this stage we have all the information we need so as to estimate
the effect of the thermal bath on the composite system AB. However, if we
want to know how is the decoherence process for the subsystem A,
we have to trace over all the degrees of freedom that belong to the
new environment. If we take a closed look at last expression, we
notice that we can assume that our new problem is a subsystem A
and a subsystem B which are coupled through an ``effective
interaction" $S_{\rm eff}^{\rm int}(x,q,x',q')$ defined by \beq
S_{\rm eff}^{\rm int}(x,q,x',q')= S_{\rm AB}(x,q)- S_{\rm AB}(x',q')
- 2 M_{\rm B} \gamma_0
\int_0^t ds \Delta q(s) \dot{\Sigma} q(s)+
i\frac{2M_{\rm B}\gamma_0k_B T}{\hbar} \int_0^t ds
(\Delta q(s))^2. \eeq

\subsection{The functional approach applied to subsystem $A$}

If we want to study the effect of subsystem B and the environment
$\cal E$ on subsystem A, we have to analyze the reduced density
matrix but for this subsystem only. That is to say, we need

\beq \rho_{\rm r}(x,x\prime)=\int_{-\infty}^{\infty}dq
\int_{-\infty}^{\infty}dq\prime \rho(x,q \vert x\prime,q \prime)
\delta(q-q\prime), \nonumber \eeq which is propagated in time by the reduced
evolution operator $\hat{{\cal J}}_{\rm r}(x,x\prime)$

\beq
\rho_{\rm r}(x,x\prime;t) = \int_{-\infty}^{\infty}dx_0
\int_{-\infty}^{\infty}dx_0\prime {\cal J}_{\rm r}(x,x\prime;t\vert
x_0,x_0\prime;0) 
~ \rho_{\rm r}(x_0,x_0\prime;0).\label{finalrho}\eeq

For simplicity we take
that at $t=0$ the subsystem A and the new environment are also
uncorrelated, i.e., $\hat{\rho}_{\rm AB}(t=0) = \hat{\rho}_{\rm A}(t=0) 
\otimes \hat{\rho}_{\rm B}(t=0)$. We assume a Gaussian wave packet of the form
 $e^{-((q_0-q_0\prime)^2)/2 \sigma}$ as the initial condition
of subsystem B . This is
a convenient choice in the sense that all of these states form a closed set
under linear evolution \cite{robin,diana}.
Then the evolution operator does not depend on the initial state of
the system and can be written as in \cite{Hu}

\beq {\cal J}_{\rm r}(x_{\rm f},x_{\rm f}\prime;t \vert x_0,x_0\prime;0)
=  \int_{x_0}^{x_{\rm f}} {\cal D}x \int_{x_0\prime}^{x_{\rm
f}\prime} {\cal D}x\prime  e^{\frac{i}{\hbar} (S_{\rm A}(x)-S_{\rm
A}(x\prime))} {\cal F}(x,x\prime) \equiv  
\int_{x_0}^{x_{\rm f}} {\cal D}x \int_{x_0\prime}^{x_{\rm f}\prime}
{\cal D}x\prime ~ \exp\left\{{\frac{i}{\hbar} {\cal
A}(x,x\prime)}\right\}, \label{calprop}\eeq 
where we have defined
${\cal F}(x,x\prime) = e^{\frac{i}{\hbar} \delta {\cal
A}(x,x\prime)}$ and ${\cal A}(x,x\prime)= S_{\rm A}(x)-S_{\rm
A}(x\prime)+\delta {\cal A}(x,x\prime)$ as the new influence
functional and influence action, respectively. In order to evaluate
$ \delta {\cal A}(x,x\prime)$, we must do the following
integrations,

\beq \delta {\cal
A}(x,x\prime) = \int_{-\infty}^{\infty} dq_0 \int_{-\infty}^{\infty} dq_0\prime
\int_{-\infty}^{\infty} dq_{\rm f} \int_{q_0}^{q_{\rm f}} {\cal D} q 
\int_{q_0'}^{q_{\rm f}'} {\cal D} q'
e^{\frac{i}{\hbar}(S_{\rm B}(q)-S_{\rm B}(q\prime))}
e^{\frac{i}{\hbar}S_{\rm eff}^{\rm int}(x,q,x',q')} .\label{deltaA2}\eeq

In order to perform the functional integrations, we must solve the classical
equation of motion for the subsystem B given by

\beq \ddot q(s)
\pm \Omega^2 q(s)=\frac{\lambda}{M_{\rm B}} x(s).\label{ec.mov.}\eeq
In the latter expression, we have neglected the term related to the
dissipation introduced by the external
environment, because we assume an underdamped environment (small $\gamma_0$).

At this stage, we must make clear that we want to analyze all
possible  combinations of subsystems A and B made up with a
harmonic oscillator and one upside-down oscillator. Then, 
in some cases, subsystem B will be a harmonic
oscillator (sign plus in Eq. (\ref{ec.mov.})) and in others will be
an upside-down oscillator (sign minus in Eq. (\ref{ec.mov.})). We
will explicitly write the solution for one case, being possible to
obtain the other solution by just replacing $\Omega$ for $i \Omega$
in the solution presented below. However, details are
presented in the Appendix.

Let's suppose subsystem B is an upside-down oscillator (case (a)) obeying
$\ddot q(s) - \Omega^2 q(s)=\frac{\lambda}{M_{\rm B}} x(s)$. In order to
find the solution to this equation, we must find the solution to the
homogeneous equation and to the particular one. After imposing
initial and final conditions $q(s=0)=q_0$ and $q(s=t)=q_{\rm f}$,
respectively, we write the complete solution as \beqa
q_{\rm cl}(s) &=&q_0\frac{\rm {\sinh}(\Omega(t-s))}{\rm
{\sinh}(\Omega t)}+ q_{\rm f} \frac{\rm {\sinh}(\Omega s)}{\rm
{\sinh}(\Omega t)} -\frac{\lambda}{M_{\rm B} \Omega}
\frac{\rm {\sinh}(\Omega s)}{\rm {\sinh}(\Omega t)} \int_0^t x(u)
\rm{\sinh}(\Omega (s-u)) du \nonumber
\\ & + & \frac{\lambda}{M_{\rm B} \Omega}\int_0^s x(u) \rm{\sinh}(\Omega
(s-u)) du. \label{classol}\eeqa

Once the full expression for $q_{\rm cl}(s)$ is known, we can go back to
Eq.(\ref{deltaA2}) and estimate it, obtaining as a result the effective
influence action for subsystem A,
\beq
\delta {\cal A}(x,x\prime)= 2 \int_0^t ds_1 \int_0^{s_1}
ds_2 ~y(s_1) \tilde{\eta}(s_1-s_2) r(s_2) + i \int_0^t ds_1
  \int_0^{s_1} ds_2 ~y(s_1) \tilde{\nu}(s_1-s_2) y(s_2), \label{newdeltaA}
\eeq with
  $y(s)=x(s)-x\prime(s)$ and $r(s)=(x(s)+x\prime(s))/2$. The
quantities $\tilde{\eta}$ and $\tilde{\nu}$ may be defined as the
new kernels of dissipation and noise, respectively, given by

\beqa \tilde{\eta}(s_1-s_2)=\frac{\lambda^2}{2 M_{\rm B} \Omega}
\rm{\sinh}(\Omega(s_1-s_2)),\nonumber \\
\tilde{\nu}(s_1-s_2)=\frac{\lambda^2\sigma}{32 \hbar}
\rm{\cosh}(\Omega(s_1-s_2)).\label{nuevanu}\eeqa

In order to evaluate this new influence functional, we will use the
saddle point method and, in this way, get rid of the functional
integrals. Since the potentials in our model are harmonic,
an exact evaluation of the path integral can be done. These integrals are
dominated by the classical solution of the free equation of motion
for subsystem A \cite{kapral1}. At this stage, we assume that our
subsystem A is a harmonic oscillator, (being possible to obtain
the solution for an upside-down oscillator by just replacing
$\omega$ for $i \omega$), obeying $\ddot x(s) + \omega^2
x(s)=0$.

If we ask for initial and final conditions of the form
$x(s=0)=x_0$ and $x(s=t)=x_{\rm f}$, the classical solution is, \beq
x_{\rm cl}(s)=x_0 \frac{\rm {\sin}(\omega(t-s))}{\rm {\sin}(\omega t)} +
x_{\rm f} \frac{\rm{\sin}(\omega s)}{\rm {\sin}(\omega t)}.\eeq

Therefore, we can write the reduced evolution operator as in Eq.(\ref{calprop}),
and the reduced density matrix following Eq.(\ref{finalrho}).

Finally, the reduced density matrix for the subsystem A takes the form,
\beq
\rho_{\rm r}(x_{\rm f},x_{\rm f}\prime;t) =
\int_{-\infty}^{\infty} dx_0 \int_{-\infty}^{\infty} dx_0\prime
\int_{x_0}^{x_{\rm f}} {\cal D}x \int_{x_0'}^{x_{\rm f}'} {\cal D}x'
~e^{\frac{i}{\hbar}U(t)} e^{-D(t)}
\rho_{\rm A}(x_0,x_0';0),\eeq
with U and D, related to the unitary evolution and decoherence
process respectively, given by,

\beq
U=(x_0-x_0\prime)\frac{\sin (\omega(t-s))}{\sin (\omega t)}
+(x_{\rm f}-x_{\rm f}\prime)\frac{\sin (\omega s)}{
\sin (\omega t)} - 2\gamma_0 \int_0^t ds_1
\int_0^{s_1}ds_2 ~ y(s_1)\tilde{\eta}(s_1-s_2)r(s_2),\eeq and

\beq D =\frac{2 \gamma_0 k_B T}{\hbar \Omega^2}
\lambda^2\int_0^t ds (\Delta q_{\rm cl}(s))^2 +
\frac{\lambda^2\sigma}{32 \hbar}\int_0^t ds_1\int_0^{s_1}ds_2
y(s_1)\tilde{\nu}(s_1-s_2)y(s_2)\label{propagator}\eeq 
From the last equation, we can see two contributions to the diffusion
coefficient. The first one, proportional to the environmental
temperature, comes from the coupling of subsystem B to the
bath. The second term is the backreaction of subsystem B over
A, through the weak $\lambda$-coupling. Even though we are working in
the high temperature limit, the underdamped bath ($\gamma_0 \ll
\Omega$) produces both contributions could be of the same order of
magnitude. Thus, both terms will be relevant in order to study
decoherence effects on subsystem A.

\section{Diffusion coefficient in the master equation}

In this Section we will derive the diffusion coefficient in the master
equation which will quantify the decoherence suffered by subsystem A
for all four different cases.
A commonly proposed way to analyze decoherence is by examining how
the non-diagonal elements of the reduced density matrix evolve under
the master equation.
Following the same techniques used for the quantum Brownian motion \cite{Hu}
to obtain the master equation we must compute the time derivative of the
propagator ${\cal J}_{\rm r}$,
and eliminate the dependence on the initial conditions $x_0$, $x_0\prime$
that enters through the classical solution $x_{\rm cl}(s)$.
This can be easily done using the properties of the solution \cite{hu2}

\beq
\Delta_0 J_{\rm r} (t,0)=\left[\cos(\omega(t-s))\Delta_{\rm f} + \frac{\sin(\omega(t-s))}
{\omega} i \hbar \frac{\partial}{\partial \Sigma_{\rm f}} \right] J_{\rm r}(t,0)
\eeq
where $\Delta_0=(x_0-x_0\prime)$, $\Delta_{\rm f}=(x_{\rm f}-x_{\rm f}
\prime)$ and $\Sigma_{\rm f}=
 (x_{\rm f} + x_{\rm f}\prime)$.

 The master equation is commonly presented as
\beqa i \hbar \dot{\rho}_{\rm r}(x,x';t)&=&\bigg[-\frac{\hbar^2}{2 M_{\rm A}} \bigg[
\frac{\partial^2}{\partial x^2}-\frac{\partial^2}{\partial x'^2}\bigg]  + 
\frac{1}{2} M_{\rm A}\Omega^2(x^2-x'^2)\bigg] \rho_{\rm r}(x,x';t) \nonumber \\ &+ &
\frac{1}{2}M_{\rm A} \delta\Omega^2(t)
(x^2-x'^2) \rho_{\rm r}(x,x';t) - 
 i \hbar \Gamma(t)(x-x')\bigg[\frac{\partial}
{\partial x}-\frac{\partial}{\partial x'}\bigg]  \rho_{\rm r}
(x,x';t) \nonumber \\ &- & i M_{\rm A} {\cal D}(t)
(x-x')^2 \rho_{\rm r}(x,x';t) - 
\hbar \Gamma(t) \textit{f}(t)(x-x')\bigg[\frac{\partial}
{\partial x}+\frac{\partial}{\partial x'}\bigg] \rho_{\rm r}(x,x';t)
\label{master},\eeqa
where ${\cal D}(t)(x-x')^2$ is the diffusion term, which produces the decay of
the off-diagonal elements. For simplicity we omitted the subindex ${\rm f}$ to
indicate the final configuration $x_{\rm f}$.
Therefore, in order to find that coefficient, we will take a closed 
look at those terms. The total diffusion coefficient is given by

\beq {\cal D}(t) =\frac{2 \gamma_0 k_B T}{\hbar \Omega^2}
\lambda^2
\int_0^t ds ~\Delta q_{\rm cl}(s)~ \dot{\Delta} q_{\rm cl}(s) 
+   \frac{\lambda^2\sigma}{32 \hbar} \int_0^t ds ~\tilde{\nu}(t-s)
~\Delta x_{\rm cl}(s) \label{D},\eeq where $\Delta q_{\rm cl}(s)$
and $\tilde{\nu}(t-s)$ are presented above in Eqs.(\ref{classol})
and (\ref{nuevanu}) respectively. It is important to note that
$\Delta q_{\rm cl}(s)$ is the solution of the coupled system, and
the new noise kernel is not the usual T-dependent noise kernel of
the quantum Brownian motion. This term is solely coming from the
interaction between the subsystems.

In the rest of this Section we will present the exact results for the
diffusion coefficients in all the four different situations considered. This
can be summarized as:

\begin{itemize}
\item {\bf Case ({\mbox a}): Harmonic Oscillator + Upside-Down Oscillator + ${\cal E}$}

This is the case we have developed so far of having a harmonic
oscillator (subsystem A) coupled to an upside-down oscillator
(subsystem B) by the interaction term presented in Eq.
(\ref{intAB}). The diffusion coefficient for this case is called
${\cal D}_{\rm a}$. This is the generalization of the toy model
considered in Ref.\cite{robin} where they did not consider the
interaction of subsystem B (upside-down
 oscillator) with an external environment. It is easy to find results of \cite{robin}
just by setting $\gamma_0 = 0$ in our results. We will plot this
case in order to compare it with the other cases. Case ({\mbox a})
is the situation in which a Brownian particle (in a harmonic
potential) suffers decoherence from an environment with one (or
more) unstable degrees of freedom.

\item {\bf Case ({\mbox b}): Upside-Down  Oscillator + Harmonic Oscillator + ${\cal E}$}

In this case we consider that subsystem A is an upside-down
oscillator, obeying the classical equation of motion $\ddot x(s) -
\omega^2 x(s)=0$ and subsystem B is a harmonic oscillator satisfying
$\ddot q(s) + \Omega^2 q(s)=\frac{\lambda}{M_{\rm B}} x(s)$. It is
straightforward to read the new solutions from the ones presented
above by replacing $\omega \rightarrow i \omega$ and $\Omega
\rightarrow i \Omega$. The diffusion coefficient for this case is
${\cal D}_{\rm b}$ (see Appendix for details), and it represents the
possibility of studying decoherence induced on an unstable system
(toy model for a chaotic subsystem) by a completely harmonic
environment \cite{diana,pau,ZP}. We will see that this is the most
decoherent system among all four cases studied in this
paper.

\item {\bf Case ({\mbox c}): Harmonic Oscillator + Harmonic Oscillator + ${\cal E}$}

For completeness, we also consider the case of two harmonic
oscillators coupled together by the interaction term (\ref{intAB})
and one of them (subsystem B) coupled to an external environment,
very hot but underdamped. The procedure for deriving the diffusion
coefficient is similar to what was done above but using the solution
of a classical harmonic oscillator for subsystems A and B, as shown
in the Appendix. It is direct to find an analytical expression for
this diffusion term (called ${\cal D}_{\rm c}$) just by replacing
$\Omega \rightarrow i \Omega$ in case (a).

\item {\bf Case ({\mbox d}): Upside-Down  Oscillator + Upside-Down Oscillator +
${\cal E}$}

Finally, we consider two  upside-down oscillators coupled together
by the same interaction term than all other cases, and one of them
coupled to ${\cal E}$. This diffusion coefficient is ${\cal D}_{\rm
d}$, and can be obtained from ${\cal D}_{\rm b}$ by making the
change $\Omega \rightarrow i \Omega$. We will see that this case is
the most sensitive to external perturbations (both subsystems are
unstable) when there is no external environment, thus decoherence is
much effective than in the other cases. In particular, it is
interesting to note that this case decoheres long before the others
when there is no thermal environment ($\gamma_0=0$).

\end{itemize}

\begin{figure}[!h]
\begin{center}
\includegraphics[width=12cm]{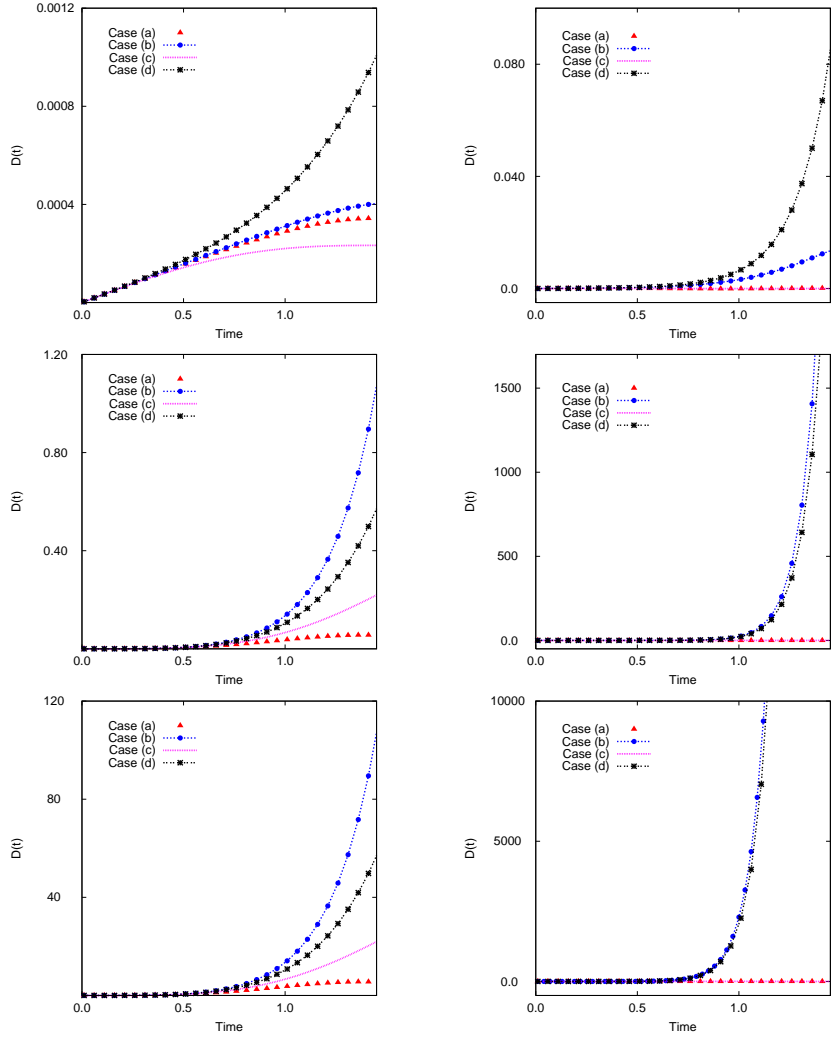}
\caption{We show the comparison between diffusion coefficients for the different
cases considered. Plots on the top refers to the case $\gamma_0=0$,
in which the composite system is isolated from the external bath.
Plots in the middle and at the bottom represent the coefficients for $\gamma_0k_BT =1$
and 100, respectively. On the left column: we use $\omega = \Omega = 1$, $\sigma =
0.01$. On the right column:  
$\omega = 5 \Omega$, $\sigma = 0.01$. 
Cases (b) and (d) show a bigger growing rate at short times for both set of parameters.}
\label{figure1}
\end{center}
\end{figure}

Once we have the analytical expression for the diffusion coefficients, we
can study their behaviour
for different relations between the parameters of the model. In order
 to illustrate some important cases, we present
 all four $\cal{D}$ coefficients when
 both frequencies of the subsystems A and B are of the same
order of magnitude ($\omega \approx \Omega$) and  when
 $\omega > \Omega$, as shown in Fig. \ref{figure1}.
Both cases are considered in the absence of external environment
$\cal{E}$ (i.e. $\gamma_0 = 0$) and for low and high $\gamma_0 k_B
T$. We will restrict our results to times in which the
no-dissipation approximation is still valid.

In Fig. \ref{figure1}, we can appreciate the
difference between the diffusion terms for composite systems when
subsystem A is unstable and when it is not. In fact, the
exponential behavior in the  (b) and (d) cases is due to the fact
that the final ``decohered'' subsystem A is unstable
\cite{diana,pau} (the upside-down oscillator). On the contrary,
cases (a) and (c) present an oscillatory behaviour since subsystem
A is a harmonic oscillator (and the solution
 $x_{\rm cl}(s)$ for the classical equation of motion has oscillatory
 functions instead of hyperbolic function as in cases (b) and (d)).
The difference between the exponential diffusion coefficients and the harmonic 
ones is particularly manifested for times $\omega t\geq 1$. For smaller times, 
all the coefficients are equivalent. But for longer times, cases (b) and (d) 
are easily dintinguishable from (a) and (c). The dynamics of the unstable 
upside-down oscillator is much in evidence after $\Omega t \geq 1$, 
where cases (b) and (d) start to differenciate each other.

The plots on top of Fig. \ref{figure1}, are a
generalization of the results obtained by Blume-Kohout and Zurek
\cite{robin} in which they considered a harmonic oscillator
coupled to an inverted oscillator as an unique environment (there
was no coupling between subsystem B and the external reservoir,
i.e. $\gamma_0=0$ in our model). Plots in the middle and at the
bottom of the same figure are for small and high values of
$\gamma_0k_B T$ since we have explicitly considered the presence
of the external hot bath $\cal{E}$, and thus, this presence
modifies the diffusion terms adding a new contribution with
respect to the one obtained in Ref. \cite{robin}.

In the case of $\gamma_0=0$, it is easy to note that ${\cal D}_{\rm
d}$ (case (d)) grows faster (like an exponential function of time),
while the other coefficients remain with a smaller amplitude. This is
important in order to evaluate decoherence times, postponed till
next Section. Instabilities inherent to the subsystem A 
exponentially enhance the diffusion originated due to the interaction with
the environment. When the B-oscillator is also unstable, we have
more exponential sensitivity to perturbations than in any other
case. However, it is important to note that it is a very toy
model, in the sense that both oscillators are unbounded from below
and therefore will develop un-physical oscillatory divergences
\cite{robin}. Therefore, we conclude that in order to have a more
precise idea of the physical consequences of having chaotic modes
into the environment, we can not neglect the interaction of these
chaotic degrees of freedom with the rest of the world, modelled
here as an infinite set of harmonic oscillators.

The diffusion suffered by the subsystem A is the direct result of
the interaction between A and B, and between the later and the
external environment ${\cal E}$ when this last one is taken into
account. The environment reacts to this interaction and the
backreaction on the subsystem is by means of diffusion and
dissipation. Let's take for example cases (b) and (d): unstable
subsystem A is coupled to a harmonic oscillator (case (b)) or to
an upside-down oscillator (case (d)). Subsystem A handles
information to B via their coupling. In case (b), as B-oscillator
is harmonic, the diffusion process is more effective. The
B-oscillator has an oscillatory behaviour in time and is able of
providing subsystem A with diffusion periodically. On the
contrary, in case (d), the B-oscillator is unstable and unbounded
from below. The stretching of its states is boundless (see next Section). 
Thus, part of the information is transferred to the reservoir, but, at large
times, the intrinsic dynamics of the upside-down oscillator makes
the provision of subsystem A with diffusion scarce and,
consequently, less effective. A more quantitative explanation will
be given in the following Section, when we give an analytical
estimation of the decoherence time for each case.

Cases (a) and (c) are slightly different. Subsystem A is harmonic
and the environment can have an unstable degree of freedom (case
(a)) or not (case (c)). It is easy to see in Figure \ref{figure1}
that the backreaction of the full environment (B
+ $\cal{E}$) on the subsystem A  by means of the diffusion process
is more effective for  cases (b) and (d) (in this order) than for
cases (c) and (a). This is due to the fact of having an inverted
oscillator as the final subsystem A. In the
middle of these figures, the ``low temperature'' limit  is shown
[we are still working in the high temperature limit, but we are
considering the underdamped case ($\gamma_0$ is small with respect
to any of the present frequencies), therefore the coefficient
$\gamma_0 k_B T$ could be a small number still preserving the hot
bath assumption]. We can see that ${\cal D}_{\rm b}$ grows
slightly faster than ${\cal D}_{\rm d}$, and both of them are
bigger than ${\cal D}_{\rm a}$ and ${\cal D}_{\rm c}$. This can be
understood by thinking in the dynamical properties of the
composite system coupled to the external bath as we mentioned
above. Oscillator B is not a good ``diffusion handler'' if it is
unstable. At times longer than $\Omega t> 1$, states in this
oscillator are spread out too much. Diffusion must go from ${\cal
E}$ to A through B (middle-environment subsystem). It is not an
effective process at short times (all the cases have a similar 
behaviour at short times). 
The dynamical behaviour of case (a) should be similar to the one
occurring in case (b). However, the difference of having an
inverted oscillator as the subsystem A or as the intermediate
subsystem B is crucial. That is reflected in the exponential grow
(or not) of the diffusion coefficient and in the decoherence times that
we will present in the following Section.

\begin{figure}[!h]
\begin{center}
\includegraphics[width=6cm]{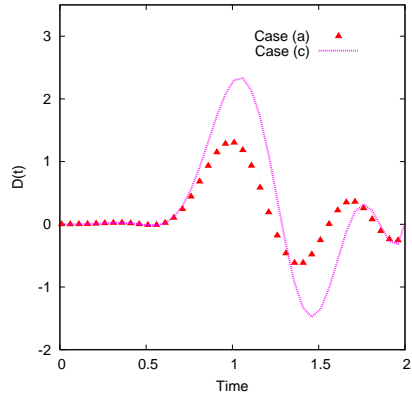}
\caption{Diffusion coefficients for cases (a) and (c) with $\omega =
5 \Omega $, $\sigma = 0.01$ and $\gamma_0 k_B T = 100$. We can see
the evolution of these coefficients at larger times than in Fig. \ref{figure1}.} 
\label{figure2}
\end{center}
\end{figure}

As we stated before, in the (b) and (d) cases, the spreading of the initial state of
the subsystem A is exponentially sensitive to fluctuations coming
from the full environment (B + $\cal{E}$), and it reacts quickly
on the bath, losing information faster than in any other case.
This is what happens in case (b), although the middle-environment
is a harmonic oscillator. However, a question might arise: why ${\cal
D}_{\rm b}$ grows faster than ${\cal D}_{\rm d}$, if the latter is
``twice'' unstable?. The key is: case (b) is the most decoherent because the
unstable system losses its information in an uniform (non-ohmic)
environment composed by a harmonic oscillator (B) plus an
infinite set of harmonic oscillators in thermal equilibrium
without internal unbounded regions (contrary
to what happens in case (d)). In the low-T example,
the major contribution to diffusion comes from the intrinsic
dynamics of the composite system alone. At early times ($\omega t
< 1$), unstable dynamics of the subsystem A dominates the temporal
behaviour and, as they both have an inverted A oscillator, both
cases (b) and (d) are similar (in fact, all the cases are similar
at short temporal scales because at very short times both
potentials are similar). However, when $\omega t \geq 1$, and the
presence of the external environment is still not so important, there is a
noteworthy difference between ${\cal D}_{\rm b}$ and ${\cal
D}_{\rm d}$ when the frequency of
the subsystem A is similar to the one of B (see middle of Fig.
\ref{figure1} on the left). When A has a bigger frequency, the dynamics is
dominated by subsystem A and both diffusion coefficients are
extremely similar in a longer temporal scale (middle Fig.
\ref{figure1} right). ${\cal D}_{\rm b}$ and ${\cal D}_{\rm d}$ are
indistinguishable for $\omega
> \Omega$ even at $\gamma_0 k_B T \sim 1$. However, as  the temperature of the 
thermal bath increases, there is no distinction between cases (b)
and (d) because the external environment dominates the diffusion 
coefficient. At high values of $\gamma_0 k_B T$ we obtain a clear
hierarchy between different composite systems (Fig. \ref{figure1}
at the bottom). Again, it is easy to observe that
the diffusion coefficients of those cases in which subsystem A is an
upside-down oscillator reach bigger values than the others that have
 a harmonic oscillator as subsystem A. Therefore, we are able to
conclude that the presence of instabilities into the composite
system enhance decoherence. This effect is yet more important if the
unstable subsystem A is coupled either to a solely chaotic degree of
freedom (in the absence of external bath like our case (d)) or to
 an external environment (formed by B + $\cal{E}$) at high $\gamma_0 k_B
 T$, as in cases (b) and (d). In Fig. \ref{figure2} we present
the diffusion coefficient at a high value of $\gamma_0 k_B T$ for
both systems which have a harmonic oscillator as subsystem A: cases
(a) and (c). This figure should come in handy so as to compare these
oscillatory coefficients with the hyperbolic-like other two.

\section{Decoherence in $A$ interacting with $B-{\cal E}$}

After integrating out all the degrees of freedom corresponding to
the external hot environment $Q_n$, and the coordinates $q$
belonging to the subsystem B, we obtained the diffusive terms that
induce decoherence on subsystem A. Therefore, we numerically
integrated the diffusive terms in time, in order to plot the
decoherence factor (see Appendix)

\beq
\Gamma (t) = \exp\left\{- \int_0^t {\cal D}(s) ~ ds\right\}.
\eeq

\begin{figure}[!h]
\begin{center}
\includegraphics[width=12cm]{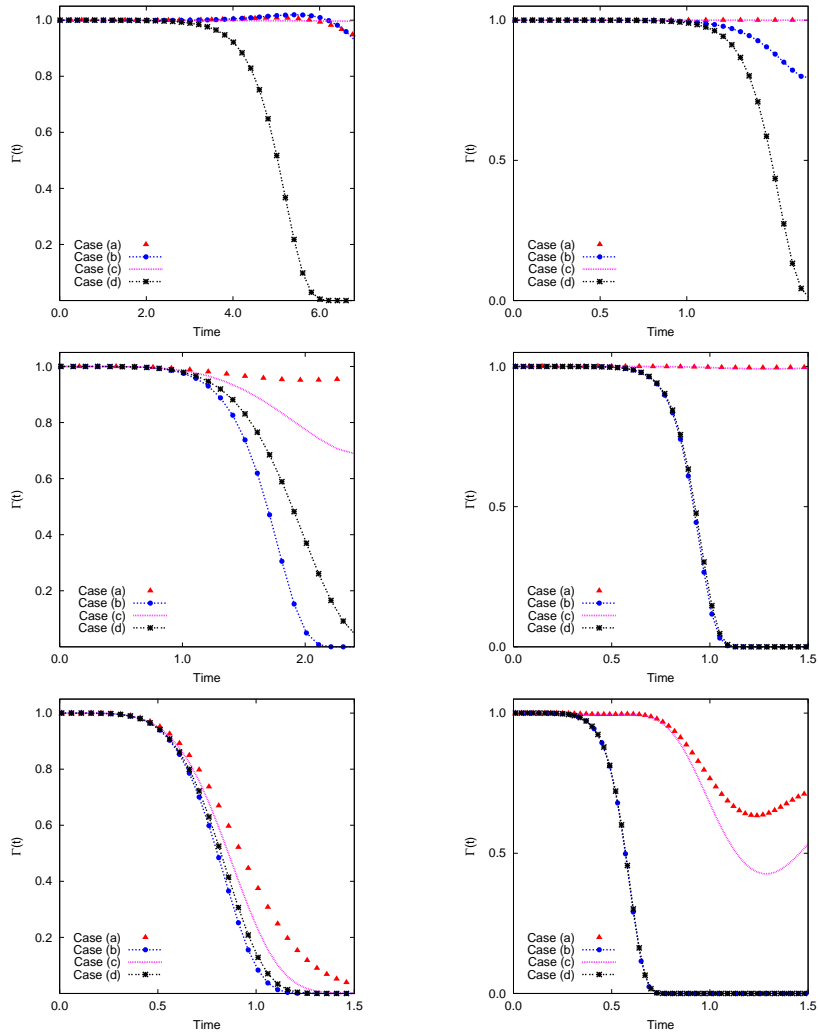}
\caption{Decoherence factor for the same set than Fig.
\ref{figure1}. Isolated composite system decoheres first for the (d)
case. For $\gamma_0\not= 0$, case (b) is more
\mbox{decoherent}.} \label{figure3}
\end{center}
\end{figure}

Thus, $\Gamma (t)$ is initially one (there is no interaction at $t
=0$ between subsystems and environment), and it decays to zero with
time (this is the case of total decoherence). From the master
equation for A-subsystem, it is easy to see that factor $\Gamma (t)$
is at the root of the loss of quantum coherence. In order to
illustrate the same cases, we present
 all four $\Gamma$ coefficients when
 both frequencies of the subsystems A and B are of the same
order of magnitude ($\omega \approx \Omega$) and  when
 $\omega > \Omega$, as shown in Fig. \ref{figure3} on the left and right columns, 
respectively. Both cases are considered in the absence of external environment
$\cal{E}$ (i.e. $\gamma_0 = 0$) and for low and high values of
$\gamma_0 k_B T$.

From the numerical results shown on top of Fig. \ref{figure3}, 
we can stress that in the absence of a hot bath, the
decoherence time is smaller in case (d) than in (b), and both of
them decohere long before cases (a) and (c). This is due to the fact
that subsystem A, which is solely coupled to subsystem B,
generates noise and dissipation at large scales. Thus,
this noise and dissipation is bigger when the subsystem B is an
upside-down oscillator (case (d)) than when it is 
a harmonic oscillator (case (b)). In this situation ($\gamma_0=0$), 
case (d) is twofold exponential in time. In these figures, we can 
also observe what is going on for cases (a) and (c). In the 
former, the oscillatory dynamics of the A-oscillator and the 
hyperbolic stretching of the B-environment, proceed largely
independently of one another. The B-environment induces only minor
perturbations in the subsystem A and this subsystem  does not
disturb the environment. The stretching of the environment (due to
being an inverted oscillator) along its unstable manifold is
reflected in the system as diffusion. The same physical process
occurs in case (b), with the sole and essential difference that
the one stretching along an unstable direction is the subsystem A,
while the environment is oscillating. As this stretching results
in diffusion, the more stretching the system has, the more
diffusion it feels. Case (d) is the best example in this
``isolated'' model, because both, A and B, stretch along a
direction in the phase space, producing double exponential diffusion.
This is the reason why it is the most decoherent case. Case (c) is
shown for completeness, but it is easily seen that decoherence
occurs in a longer time scale (there is no stretching here).

\begin{figure}[!h]
\begin{center}
\includegraphics[width=6cm]{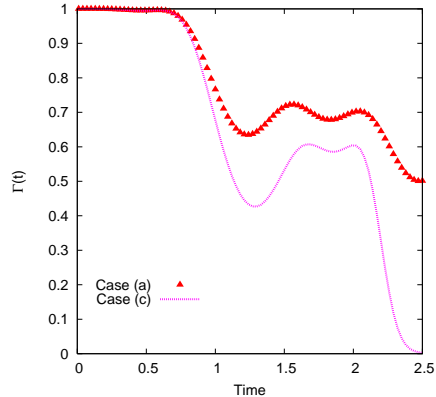}
\caption{Cases (a) and (c) for $\gamma_0k_BT = 100$, $\omega = 5
\Omega = 5$. Here we can see that, even the behaviour of
$\Gamma(t)$ is similar, case (c) decoheres first than case (a), in
which B-system is an inverted oscillator.} \label{figure4}
\end{center}
\end{figure}

As soon as the interaction between B and the thermal environment is
switched on, oscillator B dissipates not only on the bath but also
on A. This is shown in the middle and at the bottom of Fig.
\ref{figure3}. At very high environmental
temperatures, there is no difference between cases (b) and (d); both
of them decohere in the same temporal scale. The huge reservoir
dominates the diffusion terms. But they still differ from the cases
where there are harmonic oscillators as subsystems A (cases (a) and
(c)). In Figure \ref{figure4} we present the behaviour of the
$\Gamma (t)$ coefficients for these cases for a longer time scale.
We can observe that we need to wait longer times for decoherence to
be effective in cases (a) or (c) with respect to (b) and (d) even in
the highest temperature case.

\subsection{Decoherence time predictions}

In this Section we will present an analytical estimation of the 
decoherence times based on the inverted oscillators' unstable 
dynamics of the phase space. 

When the final system A is an upside-down oscillator \cite{ZP}, an
unstable point forms in the center of the phase space with
associated stable and unstable directions. These are characterized
by Lyapunov coefficients $\Lambda$ with negative  and positive
real parts respectively.  In order to have a quantitative
expression for decoherence times we have to take into account that
the dynamics now gives raise to the possibility of squeezing along
the stable direction. The exponential stretching of the Gaussian
packets in one of the directions due to the hyperbolic point is
compensated by an exponential squeezing. The time dependence of
the package width in the direction of the momenta is given by
$\sigma_p(t) = \sigma_p(t_0) \exp{[\Lambda t]}$, where
$\sigma_p(t_{0})$ is the corresponding width at the initial time. 
Lyapunov coefficient is given by the value corresponding to a linear 
potential $\Lambda = 2 \omega^2$.

Diffusion effects limit the squeezing of the Wigner function. The
bound on the width of the packets is given by $\sigma_{\rm
c}=\sqrt{2{\cal D}_{\rm i}/\Lambda}$ \cite{JPP-LH,habib} (where
${\rm i}$ is b or d). There is another scale, $t_{\rm max}$
corresponding to the time in which decoherence starts to be
effective, and after which squeezing becomes of the order of the
limiting value. We use this to estimate the decoherence time
scale.

Evolution of the Gaussian packet will typically proceed in two
different stages. During the first stage, evolution is dominated by
the unitary part of the master equation and will be within 
an approximately preserved area. This lasts the time needed for the spreading of the
initial state over a regular patch to be larger than the critical
width. During this stage diffusion does not alter much the Wigner
function, which is stretched or contracted. When the dimension of
the patch becomes comparable with $\sigma_{\rm c}$, diffusion
begins to dominate and the second stage of evolution begins.
Further contraction will be halted at $\sigma_{\rm c}$ but the
stretching will proceed at the rate set by the positive Lyapunov
exponent. As a result, the area (or the volume) in phase space
increases. One can estimate the time corresponding to the
transition from reversible to irreversible evolution as

\beq
t_{\rm c} = \frac{1}{\Lambda} \ln{\frac{\sigma_p(0)}{\sigma_{\rm c}}}.
\label{criticalt}\eeq
In our toy model, we can use this scale as the typical scale for decoherence,
setting $t_D \approx \frac{1}{\Lambda} \ln{\frac{\sigma_p(t_{\rm max})}
{\sigma_{\rm c}}}$, therefore we obtain

\beq
t_D =  t_{\rm max} + \frac{1}{\Lambda} \ln{\frac{\sigma_p(0)}{\sigma_{\rm c}}}.
\label{decotime}\eeq

For the same parameters used in Fig. \ref{figure3}, we are able 
to numerically estimate decoherence 
times as: $t_{D_{\rm b}} \sim 7.7$ and  $t_{D_{\rm d}} \sim 6.4$,
for the first set of parameters on the left of Fig. \ref{figure3} ($\omega =
\Omega = 1$) where $\gamma_0 = 0$; $t_{D_{\rm b}} \sim 2.4$ and
$t_{D_{\rm d}} \sim 2.7$; for $\gamma_0k_B T = 1$, and $t_{D_{\rm
b}} \sim 1.6$ and $t_{D_{\rm d}} \sim 1.7$, in the hight T case
$\gamma_0k_B T = 100$. For the set on the right  Fig. \ref{figure3} ($\omega
= 5\Omega = 5$), we estimated: for $\gamma_0 = 0$; $t_{D_{\rm b}}
\sim 3.0$ and $t_{D_{\rm d}} \sim 2.7$. We also got $t_{D_{\rm
b,d}} \sim 0.1$, for $\gamma_0k_B T = 1$, and $t_{D_{\rm b,d}}
\sim 0.6$ in the case $\gamma_0k_B T = 100$. All these results
agree with the decoherence times, defined by the times the
decoherence factor $\Gamma (t)$ goes to zero, that can be seen in
the figures above.

Using Eq.(\ref{criticalt}), we can check that decoherence proceeds slower in case (d) than
in case (b) for $\gamma_0 \not= 0$,

\beq t_{D_{\rm b}} - t_{D_{\rm d}} = \frac{1}{\Lambda}
\ln{\frac{\sigma_{\rm c}^{\rm d}} {\sigma_{\rm c}^{\rm b}}}=
\frac{1}{2\Lambda}\ln{\frac{{\cal D}_{\rm d}}{{\cal D}_{\rm b}}}.
\eeq By simple inspection of diffusion coefficients in the
figures, we can see that ${\cal D}_{\rm d} < {\cal D}_{\rm b}$
implying $t_{D_{\rm b}} < t_{D_{\rm d}}$.

In the isolated from external environment case ($\gamma_0=0$), we have  ${\cal
D}_{\rm d} > {\cal D}_{\rm b}$ resulting in $t_{D_{\rm b}} >
t_{D_{\rm d}}$, which agrees with our qualitative arguments and
with what is shown in the plots on top of Fig. \ref{figure3}.

Decoherence times for cases (a) and (c) occur as for the usual 
harmonic systems. We can estimate them by using the result of the
high temperature limit of the quantum Brownian motion paradigm,
i.e. $t_D$ is the solution of: $1 \approx L^2 \int_0^{t_D}{\cal
D}(s) ds$ (we have to take the typical distance $L$ as $2\sigma$,
proportional to the dispersion in position of our initial Gaussian
packet). We present, in Fig. \ref{figure4}, $\Gamma (t)$ for a
longer time scale in order to establish the corresponding
hierarchy in the  environmental decoherent effectiveness.

\section{Final remarks}

In this article we analyzed the decoherence induced by an
effective environment. The effective environment was considered to
be formed by part of a composite system and an infinite set of
harmonic oscillators. The composite system was considered to be
any of the four possible combinations made up with a harmonic and
an inverted oscillator.

Since a set of harmonic oscillators is a stable system, small
perturbations due to the state of the coupled system do not induce
exploration of a large volume of the phase space for any
oscillator. When one considers an inverted oscillator, it can
explore its volume more efficiently when it is perturbed.

When one works with a composite quantum open system in interaction
with an stable environment, it is almost probed that the composite
system (or part of it, for example its unstable degrees of
freedom) will decohere before those systems where there are
not inverted potentials (like our case (c)). Therefore, one could
have a mixture of quantum-classical dynamics for the open
composite system due to the fact that the different parts of the
global system loss coherence at different rates.

In our article we integrated out subsystem B, in order to study
the effect of having (or not) unstable degrees of freedom into the
full environment. Then we analyzed different situations and
concluded that cases (b) and (d) are the most efficient (smaller
decoherence times) at high temperatures, and (d) is the most
diffusive case, when one turns off the thermal bath. There is a
clear hierarchy between the different compositions of the
composite systems. Those in which oscillator A is unstable (cases
(b) and (d)) decohere before than those with a harmonic oscillator
as the A-subsystem (cases (a) and (c)). At high temperatures of
the external environment, it has been shown that cases (b) and (d)
have the same decoherence time scale, while composite system (c)
losses quantum coherence before case (a).

As the system and environment interact, information about the
initial state of the subsystem A is transferred to the environment
(and vice-verse). Case (b) is the most decoherent because the
unstable system losses its information in an uniform (non-ohmic)
environment composed by an harmonic oscillator (B) plus an
infinite set of harmonic oscillators in thermal equilibrium
without internal unstable (and unbounded) regions (contrary
to what happens in case (d)). At intermediate temperatures, unstable
A-subsystems decohere before stable A-subsystems because the
intrinsic dynamics of those subsystems produce exponentially
driven diffusive terms. However, at high temperatures the
environment does not distinguish between subsystems and
decoherence proceeds equally in every case. The effectiveness of
diffusion depends on the intermediate-environment subsystem B. We
have shown that harmonic oscillators keep information during a
major period of time and therefore are able to react on the
A-system more efficiently, than the case in which one has an
inverted oscillator as ``information delivery subsystem''. This is
the main reason why case (b) is more decoherent (in general) than
case (d), and why case (c) is more effective losing quantum
coherence than case (a). It is important to stress that in order 
to have a more physical model for the effective environment which 
would contain an unstable degree of freedom one should consider 
a double well potential as B-oscillator. For example, for cases 
(a) or (d). This non-linear potential has all the unstable 
properties of the upside-down oscillator at early times (giving 
a contribution to the diffusion terms which is exponential with time). 
Furthermore, this potential is bounded, which would be of much relevance 
in order to evaluate its global effect on subsystem A. In this situation, case 
(d) would be the most decoherent case at any value of $\gamma_0k_BT$ \cite{fut}.

As was emphasized by authors of Ref. \cite{robin}, the scale
associated with the decoherence process when upside-down
oscillators are taken into account as part of the coupled system,
is logarithmically dependent on the coupling constant. This is
easy to see from our analytical results, using the expression of
the diffusion coefficient ${\cal D}$, given in the Appendix
(Eq.(\ref{D})), into Eq.(\ref{decotime}). This implies that
isolation from chaotic environments is ``exponentially''
difficult. It is even harder to isolate a system (or subsystem)
from a chaotic environment than from the many harmonic oscillators
of the quantum Brownian motion environment, where decoherence time
is quadratic in the coupling constant.

\section{Acknowledgments}
We thank Juan Pablo Paz for useful discussions. We also thank F.D. Mazzitelli and
W.H. Zurek for comments during the early stage of this work.  This work was
supported by UBA, CONICET, Fundaci\'on Antorchas, and ANPCyT,
Argentina.

\begin{appendix}
\section{Derivation of the diffusion coefficient}

In this Section, we show the calculation of diffusion coefficient
corresponding to case (a), in which we have a harmonic oscillator
coupled to an upside-down oscillator, and it is in interaction
with a set of infinite harmonic oscillators at temperature T.

In order to perform the functional integrations of
Eq.(\ref{deltaA2}), we must solve the classical equation of motion
for the subsystem B. If this system satisfies,
$\ddot q(s) - \Omega^2 q(s)=\frac{\lambda}{M_{\rm B}} x(s)$.
The complete solution to this equation, after imposing initial
 and final conditions $q(s=0)=q_0$ and $q(s=t)=q_{\rm f}$, is given
in Eq.(\ref{classol}).

Once we have the classical solution for the
upside-down oscillator coupled to a subsystem of coordinate $x(s)$,
we can write explicitly the influence action obtained after
integrating out all the degrees of freedom of the environment $\cal
E$ (Eq.(\ref{deltaA})). Neglecting the action of dissipation into
the influence action (the dissipation coefficient $\gamma_0$
is basically the square of the coupling constant between subsystem
$B$ and the environment ${\cal E}$,
[$\gamma_0 \approx c_n^2$] and we are working in the underdamped
and high temperature limit ($\gamma_0 < \omega, \Omega < k_B T$)),
we can write it as

\beq \delta A(q,q\prime)=
 i 2 M_{\rm B} \gamma_0 K T \int_0^t ds(\Delta q(s))^2. \label{deltaapend}\eeq
In last equation, $\Delta q(s)$ is given by

\beq \Delta q_{\rm cl}(s)= (q_0
-q_0\prime)^2 \bigg(\frac{\sinh(\Omega(t-s))}{\sinh(\Omega
t)}\bigg)^2 + 2(q_0-q_0\prime)
\frac{\sinh(\Omega(t-s))}{\sinh(\Omega t)} g(s,t)+ g(s,t)^2,
 \eeq
with

\beq g(s,t)= \frac{\lambda}{M_{\rm B} \Omega}
\bigg(-\int_0^s du \Delta x(u) \sinh(\Omega(s-u)) 
+ \frac{\sinh(\Omega s)}{\sinh(\Omega t)} \int_0^t du \Delta
x(u) \sinh(\Omega(t-u))\bigg)\label{g} \eeq and $\Delta
x(u)=x(u)-x\prime(u)$. As the last term in expression (\ref{g})
does not depend on the initial conditions, it will be transparent
to integrals in Eq. (\ref{deltaA2}). It is important to note that
as subsystem $B$ is an upside-down oscillator with frequency
$\Omega$, all the time-dependent functions in Eq. (\ref{g}) are
hyperbolic trigonometric functions.

After imposing $\rho_{\rm B}(q_0,q_0\prime;0)= N e^{-((q_0-q_0\prime)^2)/2
\sigma}$ as the initial wave packet for subsystem $B$, the
expression for the influence functional is
\beqa
{\cal F}(x,x') &=&  \exp\left\{{{i\over{\hbar}}
\delta A(q,q')}\right\}=
\int_{-\infty}^{+\infty}dq_{\rm f}
\int_{-\infty}^{+\infty}dq_0\int_{-\infty}^{+\infty} dq_{\rm f}'
\int_{q_0}^{q_{\rm f}}{\cal D}q\int_{q_0'}^{q_{\rm f}'}{\cal D}q'
e^{\frac{i}{\hbar} \delta A(q_{\rm cl},q_{\rm
cl}\prime)} \ \rho_{\rm r}^{\rm B}(q_0,q_0') \nonumber \\
&\times & e^{{i\over{\hbar}}
(S_{\rm B}(q_{\rm cl}) + S_{\rm AB}(x,q_{\rm cl}) -
S_{\rm B}(q'_{\rm cl}) - S_{\rm AB}(x',q'_{\rm cl}))} 
 ,\eeqa where
$q_{\rm cl}(s)$ is given in Eq.(\ref{classol}) and $\delta
A(q,q\prime)$ in Eq. (\ref{deltaapend}). Therefore, this
integration can be done and yields (for a weakly coupled composite
AB system) the result shown in Eqs.(\ref{newdeltaA}) and
(\ref{nuevanu}).

Once the influence functional is known, it is straightforward to
write down the reduced density matrix for subsystem $A$ only

\beq
\rho_{\rm r}(x_{\rm f},x_{\rm f}\prime;t) = 
\int_{-\infty}^{\infty} dx_0 \int_{-\infty}^{\infty} dx_0\prime
\int_{x_0}^{x_{\rm f}} {\cal D}x \int_{x_0'}^{x_{\rm f}'} {\cal D}x'
e^{\frac{i}{\hbar} (S_{\rm A}(x)-S_{\rm A}(x\prime))} 
e^{-g^2(s,t)} e^{\frac{i}{\hbar} \delta {\cal
A}(x,x\prime)} \rho_{\rm A}(x_0,x_0\prime,0),\eeq
and the reduced evolution operator

\beq
{\cal J}_{\rm r}(x_{\rm f},x_{\rm f}\prime;t\vert x_0,x_0\prime;0)
= \int_{x_0}^{x_{\rm f}}{\cal D}x \int_{x_0\prime}^{x_{\rm f}\prime} {\cal D} x
\prime e^{\frac{i}{\hbar}(S_{\rm A}(x)-S_{\rm A}(x\prime))} 
e^{-g^2(s,t)} e^{\frac{i}{\hbar} \delta {\cal
A}(x,x\prime)}\label{Jr}.\eeq

In order to evaluate Eq.(\ref{Jr}), we need the solution to the free equation
of motion for subsystem A $\ddot x(s) + \omega^2 x(s)=0$. 
If we ask for initial and final conditions of the form $x(s=0)=x_0$ and
$x(s=t)=x_{\rm f}$, the classical solution is, $x_{\rm cl}(s)=x_0
\frac{\rm {\sin}(\omega(t-s))}{\rm {\sin}(\omega t)} + x_{\rm f}
\frac{\rm{\sin}(\omega s)}{\rm {\sin}(\omega t)}$,
and the reduced evolution operator becomes

\beq 
{\cal J}_{\rm r}(x_{\rm f},x_{\rm f}\prime;t\vert x_0,x_0\prime;0)
=  e^{\frac{i}{\hbar}(S_{\rm A}(x_{\rm cl})-S_{\rm A}(x\prime_{\rm cl}))}
e^{-g^2(s,t)}
e^{\frac{i}{\hbar} \delta {\cal A}(x,x\prime)} \equiv ~e^{\frac{i}{\hbar}U(t)} e^{-D(t)}
\label{Jr2}.\eeq \\
with U and D, related to the unitary evolution and decoherence
process respectively, given by,
\beq
U=(x_0-x_0\prime)\frac{\rm {\sin}(\omega(t-s))}{\rm
{\sin}(\omega t)}+(x_{\rm f}-x_{\rm f}\prime)\frac{\rm {\sin}(\omega s)}{\rm
{\sin}(\omega t)} - 2\gamma_0 \int_0^t ds_1
\int_0^{s_1}ds_2  y(s_1)\tilde{\eta}(s_1-s_2)r(s_2),\nonumber \eeq
and

\beq
D=\frac{2 \gamma_0 k_B T}{\hbar\Omega^2} \lambda^2 \int_0^t ds (\Delta
q(s))^2 +   \frac{\lambda^2}{32 \hbar \sigma}\int_0^t ds_1\int_0^{s_1}ds_2
y(s_1)\tilde{\nu}(s_1-s_2)y(s_2).\nonumber\eeq

Following the same techniques used for the quantum Brownian motion
\cite{Hu} to obtain the master equation, we must compute the time
derivative of the propagator ${\cal J}_{\rm r}$ Eq.(\ref{Jr2}),
and eliminate the dependence on the initial conditions
$x_0$, $x_0\prime$ that enters through the classical solution
$x_{\rm cl}(s)$. This can be easily done using the properties of
the solution \beq \Delta_0 J_{\rm
r}(t,0)=\left[\cos(\omega(t-s))\Delta_{\rm f} + \frac{\sin(\omega(t-s))}
{\omega} i \hbar \frac{\partial}{\partial \Sigma_{\rm f}} \right] J_{\rm r}(t,0)
\eeq
 where $\Delta_0=(x_0-x_0\prime)$, $\Delta_f=(x_f-x_f\prime)$ and $\Sigma_f=
 (x_{\rm f}+x_{\rm f}\prime)$.
This identity allows us to remove the initial coordinates $x_0$, $x_0'$
by expressing them in terms of the final coordinates $x_{\rm f}$, $x_{\rm f}'$ and
the derivatives $\partial_{x_{\rm f}},\partial_{x_{\rm f}'}$, and obtain the
master equation.

The full equation is very complicated and, like in the case of the
quantum Brownian motion, it depends on the system-environment
coupling. In the present case, it also depends on the subsystems
coupling constant $\lambda$. As we are solely interested in
decoherence, it is sufficient to calculate the correction to the
usual unitary evolution coming from the noise kernel (imaginary
part of the influence action). The result reads
 \beqa
\dot{\rho}_{\rm r}(x_{\rm f},x_{\rm f}';t) &\sim &  - i [H_{\rm ren},\rho_{\rm r}] - 
\frac{\partial}{\partial t} \bigg( \frac{2 \gamma_0 k_B
T}{\hbar}\frac{\lambda^2}{\Omega^2}  \int_0^t ds ~g(s)^2 
-  \int_0^t \int_0^s ds ds' \Delta x_{\rm cl}(s) \tilde{\nu}(s,s')
\Delta x_{\rm cl}(s')\bigg) \rho_{\rm r} + ... \nonumber \\ &= & - i [H_{\rm ren},\rho_{\rm r}] 
- \bigg( \frac{2
\gamma_0 K T}{\hbar}\frac{\lambda^2}{\Omega^2} \int_0^t 2
\dot{g}(s) g(s) ds  + \frac{\lambda^2\sigma}
{32 \hbar} (x_{\rm f}-x_{\rm f}')\int_0^t ds \cosh(\Omega(t-s))\Delta
x_{\rm cl}(s)\bigg)\rho_{\rm r} \nonumber \\
&+& ...  ,\nonumber \eeqa where ellipsis denote other terms that
do not contribute to decoherence.

This is equivalent to write \beq \dot{\rho_{\rm r}} \approx  - i [H_{\rm ren},\rho_{\rm r}] 
-(x_{\rm f}-x_{\rm f}')^2
{\cal D}(t) \rho_{\rm r}, \eeq
with $\cal{D}$ the diffusion coefficient
(Eq.(\ref{D})).  Then, the effect of the diffusion coefficient on
the decoherence process can be seen considering the following
approximate solution to the master equation

\beq \rho_{\rm r}(x_{\rm f},x_{\rm f}';t)
\approx \rho_{\rm r}^{\rm u}(x_{\rm f},x_{\rm f}';t) ~ 
e^{-(x_{\rm f} - x_{\rm f}')^2\int_0^t{\cal D}(s)ds},
\label{decofactor}\eeq
where $\rho_{\rm r}^{\rm u}$ is the solution to the unitary part of the
master equation ( i.e., without environment). The system will
decohere when the non-diagonal elements of the reduced density
matrix are much smaller than the diagonal ones.

For the case of a harmonic oscillator coupled to an upside-down
oscillator, which is coupled to an external environment, the diffusion
coefficient (Eq.(\ref{D})) for total the reduced density matrix
$\rho_{\rm r}(x_{\rm f},x_{\rm f}';t)$ is

\beqa {\cal D}_{\rm a}
&=&  \frac{\Omega^2}{(\omega^2+\Omega^2)^2}\Big\{\frac{2 \gamma_0 k_B
T}{\hbar \Omega^2} \lambda^2 \int_0^t ds 
\Big[\frac{
\sinh(\Omega s)}{\sinh(\Omega t)}(\cosh(\Omega t)\cos(\omega
t)-1)- \cosh(\Omega s)\cos(\omega t)
+\cos(\omega(t-s))\Big]\nonumber \\
&\times & \Big[\Omega\Big(\frac{
\sinh(\Omega s)}{\sinh(\Omega t)^2}\cosh(\Omega t) 
(1- \cosh(\Omega t)\cos(\omega t))+\sinh(\Omega s)\cosh(\omega t)\Big)
+  \omega \Big(-\frac{ \sinh(\Omega s)}{\sinh(\Omega
t)}\sin(\omega t)\cosh(\Omega t)\nonumber \\ &-& 
\sin(\omega(t-s)) +  \sin(\omega t)\cosh(\Omega s)\Big)\Big] + 
\frac{\lambda^2\sigma}{32 \hbar} \int_0^t
ds\cosh(\Omega(t-s))\cos(\omega(t-s)) \Big\}.\label{D}
\eeqa

The procedure to evaluate the diffusion term for case (b), when
one has an upside-down oscillator coupled to a harmonic one, and
the last to the environment, is equivalent to what was done in the
Subsection above but replacing $\Omega \rightarrow i \Omega$ and
$\omega \rightarrow i \omega$. This is because subsystem $B$
satisfies ${\ddot q}(s)+\Omega^2 q(s)=\frac{\lambda}{M_{\rm
B}} x(s)$ and subsystem $A$ satisfies ${\ddot
x}(s)-\omega^2 x(s)=0$. 

Therefore, in this case, $q_{\rm cl}(s)$ is,
\beqa
q_{\rm cl}(s)&=&q_0\frac{\rm {\sin}(\Omega(t-s))}{\rm
{\sin}(\Omega t)}+ q_{\rm f} \frac{\rm {\sin}(\Omega s)}{\rm
{\sin}(\Omega t)}- \frac{\lambda}{M_{\rm B} \Omega}
\frac{\rm {\sin}(\Omega s)}{\rm {\sin}(\Omega t)} \int_0^t  x(u)
\rm{\sin}(\Omega (s-u)) du  \nonumber
\\ &+ &  \frac{\lambda}{M_B \Omega}\int_0^s x(u) \rm{\sin}(\Omega
(s-u)) du,\label{qclb} \eeqa
and $x_{\rm cl}(s)$,

\beq x_{\rm cl}(s)=x_0
\frac{{\sinh}(\omega(t-s))}{{\sinh}(\omega t)} + x_{\rm f}
\frac{{\sinh}(\omega s)}{{\sinh}(\omega t)}.\label{xclb}\eeq

If one follows all the  procedure detailed above one obtains as the diffusion coefficient,

\beqa {\cal D}_{\rm b}
&=&  \frac{\Omega^2}{(\omega^2+\Omega^2)^2}\Big\{\frac{2 \gamma_0 k_B
T}{\hbar \Omega^2} \lambda^2 \int_0^t ds 
\Big[\frac{
\sin(\Omega s)}{\sin(\Omega t)}(\cos(\Omega t)\cosh(\omega
t)-1)- \cos(\Omega s)\cosh(\omega t)
+\cosh(\omega(t-s))\Big]\nonumber \\
&\times & \Big[\Omega\Big(\frac{
\sin(\Omega s)}{\sin(\Omega t)^2}\cos(\Omega t) 
(1- \cos(\Omega t)\cosh(\omega t))+\sin(\Omega s)\cos(\omega t)\Big)
+  \omega \Big(-\frac{ \sin(\Omega s)}{\sin(\Omega
t)}\sinh(\omega t)\cos(\Omega t)\nonumber \\ &-& 
\sinh(\omega(t-s)) +  \sinh(\omega t)\cos(\Omega s)\Big)\Big] + 
\frac{\lambda^2\sigma}{32 \hbar} \int_0^t
ds\cos(\Omega(t-s))\cosh(\omega(t-s)) \Big\}.\label{D}
\eeqa

Diffusion terms ${\cal D}_{\rm c}$ and ${\cal D}_{\rm d}$ can be
obtained in a similar way, by replacing $\Omega \rightarrow i
\Omega$ in ${\cal D}_{\rm a}$ and ${\cal D}_{\rm b}$,
respectively.

\end{appendix}

\end{document}